\documentclass[showpacs,superscriptaddress,floatfix,pra,twocolumn,10pt]{revtex4}
\usepackage{graphicx}
\usepackage{amsmath,amssymb}
\usepackage{bm}

\begin{document}

\title{Entanglement of nanomechanical oscillators and two-mode fields
induced by atomic coherence }
\author{Ling Zhou}
\affiliation{School of physics and optoelectronic technology, Dalian University of
Technology, Dalian 116024, P.R.China}
\author{Yan Han}
\affiliation{School of physics and optoelectronic technology, Dalian University of
Technology, Dalian 116024, P.R.China}
\author{Jietai Jing}
\affiliation{State Key Laboratory of Precision Spectroscopy, Department of Physics, East
China Normal University, Shanghai 200062, P.R. China}
\author{Weiping Zhang}
\affiliation{State Key Laboratory of Precision Spectroscopy, Department of Physics, East
China Normal University, Shanghai 200062, P.R. China}

\begin{abstract}
We propose a scheme via three-level cascade atoms to entangle two
optomechanical oscillators as well as two-mode fields. We show that two
movable mirrors and two-mode fields can be entangled even for bad cavity
limit. We also study entanglement of the output two-mode fields in frequency
domain. The results show that the frequency of the mirror oscillation and
the injected atomic coherence affect the output entanglement of the two-mode
fields.
\end{abstract}

\maketitle

\section{\protect\bigskip Introduction}

Entanglement among mesoscopic even macroscopic systems have always been an
attractive topic since the birth of quantum mechanics. Schr\"{o}dinger's cat
paradox is a well known example of it (entanglement between macroscopic cat
state and microscopic state). Cavity optomechanical system is an important
candidate for the study of quantum mechanical features at mesoscopic even
macroscopic scales. Recently, there has been considerable interest in
studying entanglement in mesoscopic systems [1-9]. Refs. \cite{vitali,Genes}
investigated the optomechanical entanglement between the field of an optical
cavity and a vibrating cavity end-mirror. Many proposes were put forward to
entangling two mirrors in a ring cavity \cite{Mancini}, entangling two
mirrors of two independent optical cavities driven by a pair of entangled
light beams \cite{zhangj}, entangling two mirrors of a linear cavity driven
by a classical laser field \cite{DAVID}, entangling two separated
nanomechanical oscillators by injecting broad band squeezed vacuum light and
laser light into the ring cavity \cite{huangshumei}. Most recently, Ref. %
\cite{simon} observed strong coupling between a micromechanical resonator
and an optical cavity field, which could push forward the possibility of
entangling two mirrors via optical cavity field.

On the other hand, atomic medium play important role in the interaction of
cavity electrodynamics. When the atomic medium is trapped in a cavity with
micromechanical oscillator, some interesting phenomena have been revealed %
\cite{genes2,sunatom,single,single2}. Genes \textit{et} \textit{al} \cite%
{genes2} suggested cooling ground-state of micromechanical oscillators by
resonant coupling of the mirror vibrations to a two-level atomic bath. Ian %
\cite{sunatom} found that two-level atoms effectively enhance the radiation
pressure of the cavity field upon the oscillating mirror. Most recently,
Hammerer \textit{et} \textit{al }\cite{single} and Wallquist \cite{single2}
have shown the possibility to achieve strong cavity-mediated coupling
between a single microscopic atom and a macroscopic mechanical oscillator.
In addition, Genes \cite{genes3} studied tripartite entanglement among atom,
mirror and cavity field. As we all know, atomic coherence results in many
interesting phenomena \cite{scully} such as electromagnetically induced
transparency (EIT),\ correlation emission laser (CEL), laser without
inversion (LWI). A lot of works have been done on the entanglement directly
induced by atomic coherence \cite{han,tan,zhou1,zhoupla,zhou3}. When one of
the mirrors of the cavity is movable, the atomic coherence effects are not
studied before. In this paper, we propose a method to entangle two
macroscopic mirrors via microscopic atomic coherence. When atomic beam with
cascade configuration is injected into the two-mode cavity, the two-mode
fields as well as the optomechanical oscillator are entangled. In Refs. \cite%
{single,single2}, authors treat the cavity field as a quantum bus and give
an effective coupling between oscillator and the single atomic motion.
Instead, we directly treat hybrid system. Our study show that the initial
atomic coherence and the frequency of the mirror's motion affect the
entanglement of the output fields.

\section{Model and the Hamiltonian of the system}

\bigskip The system under study is a two mode cavity with one fixed
partially transmitting mirror and two movable perfectly reflecting mirrors,
sketched in Fig. 1. The atomic medium with cascade configuration is injected
into the cavity and interacts with two-mode cavity fields with detuning $%
\Delta _{i}$, respectively ($i=1,2$). The Hamiltonian of the hybrid system
reads
\begin{eqnarray}
H &=&\sum_{j=1,2}\hslash \omega _{j}a_{j}^{\dagger }a_{j}+\sum_{j=1,2}\frac{%
\hslash \omega _{m_{j}}}{2}({P_{j}^{2}}+{Q_{j}^{2})} \\
&&+\sum_{j=1,2}i\hslash \varepsilon _{j}(a_{j}^{\dagger }e^{-i\omega
_{L_{j}t}}-a_{j}e^{i\omega _{L_{j}t}})+\sum_{j=1,2}\hslash \chi
_{j}Q_{j}a_{j}^{\dagger }a_{j}  \notag \\
&&+\sum_{i=a,b,c}\hslash E_{i}\sigma _{ii}+\hslash (g_{1}{\sigma }_{ba}{a_{1}^{\dagger }}+g_{2}{\sigma }_{cb}{%
a_{2}^{\dagger }}+h.c.)  \notag
\end{eqnarray}%
The first term describes the energy of the two cavity modes, with lowering
operator $a_{j}$, cavity frequency $\omega _{j}$, and the decay rate $\kappa
_{j}$. The second term represents the energy of the two mechanical
oscillators at frequency $\omega _{m_{j}}$,\ and $P_{j}$\ and $Q_{j}$ are
their position and momentum operators. The third term describe the two
driving laser with frequency $\omega _{L_{1}}$ and $\omega _{L_{2}}$,
respectively. The forth is the radiation-pressure coupling with rate $\chi
_{j}=\frac{\omega _{j}}{L}\sqrt{\frac{\hbar }{m\omega _{m_{j}}}}$, and last
summation describes the energy of the atoms, and the last term is
interaction between the atom and the cavity fields, where $\sigma
_{ij}=|i\rangle \langle j|$ is the spin operator of the atom.\
\begin{figure}[b]
\includegraphics[width=\columnwidth]{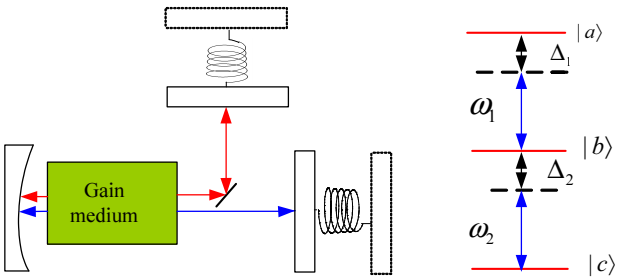}
\caption{The sketch of the system and the atomic configuration.}
\label{Fig1}
\end{figure}

In interaction picture, we have the Hamiltonian\ \ \ \ \
\begin{eqnarray}
H_{I} &=&\hslash \delta _{1}\sigma _{aa}-\hslash \delta _{2}\sigma _{cc} \\
&&+\hslash (g_{1}{\sigma }_{ba}{a_{1}^{\dagger }}+g_{2}{\sigma }_{cb}{%
a_{2}^{\dagger }}+h.c.)  \notag \\
&&+\hslash (\delta _{1}-\Delta _{1})a_{1}^{\dagger }a_{1}+\hslash (\delta
_{2}-\Delta _{2})a_{2}^{\dagger }a_{2}  \notag \\
&&+i\varepsilon _{1}\hslash (a_{1}^{\dagger }-a_{1})+i\varepsilon
_{2}\hslash (a_{2}^{\dagger }-a_{2})  \notag \\
&&+\frac{\hslash \omega _{m_{1}}}{2}({P_{1}^{2}}+{Q_{1}^{2}})+\frac{\hslash
\omega _{m_{2}}}{2}({P_{2}^{2}}+{Q_{2}^{2}})  \notag \\
&&+\hslash \chi _{1}Q_{1}a_{1}^{\dagger }a_{1}+\hslash \chi
_{2}Q_{2}a_{2}^{\dagger }a_{2}],  \notag
\end{eqnarray}%
where $\Delta _{1}=E_{a}-E_{b}-\omega _{1}$, $\Delta _{2}=E_{b}-E_{c}-\omega
_{2}$, $\delta _{1}=E_{a}-E_{b}-\omega _{L_{1}}$, $\delta
_{2}=E_{b}-E_{c}-\omega _{L_{2}}$. Thus $\delta _{j}-\Delta _{j}=\omega
_{j}-\omega _{L_{j}}$ means the detuning between the classical driving
fields and the cavity field. The dynamics of the system is determined by the
following quantum Langevin equations
\begin{eqnarray}
\dot{Q_{1}} &=&\omega _{m_{1}}P_{1},  \notag \\
\dot{Q_{2}} &=&\omega _{m_{2}}P_{2},  \label{langevin} \\
\dot{P_{1}} &=&-\chi _{1}a_{1}^{\dagger }a_{1}-\omega _{m_{1}}Q_{1}-\gamma
_{m}P_{1}+\xi _{1},  \notag \\
\dot{P_{2}} &=&-\chi _{2}a_{2}^{\dagger }a_{2}-\omega _{m_{2}}Q_{2}-\gamma
_{m}P_{2}+\xi _{2},  \notag \\
\dot{a_{1}} &=&-[\kappa _{1}+i\widetilde{\Delta }_{1}]a_{1}-ig_{1}\sigma
_{ba}+\varepsilon _{1}+\sqrt{2\kappa _{1}}a_{1in},  \notag \\
\dot{a_{2}} &=&-[\kappa _{2}+i\widetilde{\Delta }_{2}]a_{2}-ig_{2}\sigma
_{cb}+\varepsilon _{2}+\sqrt{2\kappa _{2}}a_{2in},  \notag \\
\dot{\sigma}_{ba} &=&-(\gamma +i\delta _{1})\sigma _{ba}-ig_{1}a_{1}(\sigma
_{bb}-\sigma _{aa})+ig_{2}a_{2}^{\dagger }\sigma _{ca},  \notag \\
\dot{\sigma}_{cb} &=&-(\gamma +i\delta _{2})\sigma
_{cb}-ig_{1}a_{1}^{\dagger }\sigma _{ca}-ig_{2}a_{2}(\sigma _{cc}-\sigma
_{bb}),  \notag
\end{eqnarray}%
where $\widetilde{\Delta }_{j}=\delta _{j}-\Delta _{j}+\chi _{j}Q_{j}$ ($%
j=1,2$). The quantum Brownian noise $\xi _{1}$ and $\xi _{2}$ are from the
coupling of the movable mirrors to their own environment. They are mutually
independent with zero mean values and have the following correlation
function at temperature $T$
\begin{eqnarray}
\langle \xi _{j}(t)\xi _{k}(t^{\prime })\rangle  &=&\frac{\delta _{jk}\gamma
_{m}}{\omega _{m}}\int \frac{d\omega }{2\pi }e^{-i\omega (t-t^{\prime
})}\omega \lbrack 1+\coth (\frac{\hslash \omega }{2\kappa _{B}T})],  \notag
\\
j,k &=&1,2.
\end{eqnarray}%
The two cavity modes decay at the rate $\kappa _{1}$ and $\kappa _{2}$, and $%
a_{1in}$ ($a_{2in}$) is the vacuum radiation input noise, whose correlation
functions are given by
\begin{eqnarray}
\langle a_{jin}^{\dagger }(t)a_{jin}(t^{\prime })\rangle  &=&N\delta
(t-t^{\prime }), \\
\langle a_{jin}(t)a_{jin}^{\dagger }(t^{\prime })\rangle  &=&(N+1)\delta
(t-t^{\prime }),  \notag
\end{eqnarray}%
where $N=[exp(\hslash \omega _{c}/k_{B}T)-1]^{-1}$.

In order to obtain the steady state solution of Eq.(\ref{langevin}), we
calculate the last two equations to the first order in $g_{i}$ $(i=1,2)$,
i.e., using linear approximation theory \cite{scully}, which means that in
the last equation of (\ref{langevin}), for the terms that the $\sigma _{ij}$
multiply $a$ ($a^{\dagger }$), we use zero order $\langle \sigma
_{ij}\rangle $ substitute of $\sigma _{ij}$. We assume that the atoms are
injected into the cavity with the state $\rho _{a}=\rho _{aa}^{0}|a\rangle
\langle a|+\rho _{cc}^{0}|c\rangle \langle c|+\rho _{ca}^{0}(|c\rangle
\langle a|+|a\rangle \langle c|)$ at injection rate $r_{a}$. The last two
equation of (\ref{langevin}) can be rewritten as%
\begin{eqnarray}
\dot{\sigma}_{ba} &=&-(\gamma +i\delta _{1})\sigma _{ba}+ig_{1}r_{a}{\rho
_{aa}^{0}}a_{1}+ig_{2}r_{a}{\rho _{ca}^{0}a_{2}^{\dagger },}  \label{sigma}
\\
\dot{\sigma}_{cb} &=&-(\gamma +i\delta _{2})\sigma _{cb}-ig_{1}r_{a}{\rho
_{ca}^{0}}a_{1}^{\dagger }-ig_{2}r_{a}{\rho _{cc}^{0}}a_{2}.  \notag
\end{eqnarray}%
By combining it with Eq.(\ref{langevin}), we finally have the steady-state
mean values of the system as
\begin{eqnarray}
P_{1}^{s} &=&0,P_{2}^{s}=0  \notag \\
Q_{1}^{s} &=&\frac{-\chi _{1}|a_{1}^{s}|^{2}}{\omega _{m_{1}}},Q_{2}^{s}=%
\frac{-\chi _{2}|a_{2}^{s}|^{2}}{\omega _{m_{2}}},  \label{steady} \\
a_{1}^{s} &=&\frac{s_{2c}^{\ast }\varepsilon _{1}+\varepsilon _{2}^{\ast
}u_{1}}{u_{1}u_{2}^{\ast }+s_{1a}s_{2c}^{\ast }},  \notag \\
a_{2}^{s} &=&\frac{s_{1a}^{\ast }\varepsilon _{2}-\varepsilon _{1}^{\ast
}u_{2}}{u_{1}^{\ast }u_{2}+s_{2c}s_{1a}^{\ast }}  \notag
\end{eqnarray}%
with%
\begin{eqnarray}
u_{l} &=&\frac{g_{1}g_{2}r_{a}\rho _{ca}^{(0)}}{\gamma +i\delta _{l}},l=1,2
\notag \\
s_{1a} &=&\kappa _{1}+i\widetilde{\Delta }_{1}-\frac{g_{1}^{2}r_{a}\rho
_{aa}^{(0)}}{\gamma +i\delta _{1}},  \notag \\
s_{2c} &=&\kappa _{2}+i\widetilde{\Delta }_{2}+\frac{g_{2}^{2}r_{a}\rho
_{cc}^{(0)}}{\gamma +i\delta _{2}}.
\end{eqnarray}%
\ From the expressions of (7) and (8), we see that if $g_{1}=g_{2}=0$ (no
atoms within the cavity), the steady value of $a_{j}^{s}$ ($j=1,2$) will
have the same form with \cite{Genes,huangshumei}, i.e., $a_{j}^{s}=\frac{%
\varepsilon _{j}^{\ast }}{\kappa _{j}+i\widetilde{\Delta }_{j}}$. If the
injected atoms have no coherence between their levels, that is to say, $\rho
_{ca}^{(0)}=0$ ($u_{l}=0$). By writing the last terms of $s_{1a}$ and $%
s_{2c} $ into real and imaginary parts, we know that the existence of atomic
medium only affect the effective decay rate of the photon and the radiation
pressure of the cavity field upon the two mirrors. However, if the injected
atoms have coherence between the level $|a\rangle $ and $|c\rangle $, i.e., $%
\rho _{ca}^{(0)}\neq 0$, the two mode fields will be correlated so that the
two mirrors will be dependent each other, see (7) and (8).

We write each operators of the system as the sum of their steady-state mean
value and a small fluctuation with zero mean value. The fluctuations can be
calculated analytically by using the linearization approach of quantum
optics \cite{huangshumei,scully,out}, provided that the nonlinear effect
between the cavity field and the movable mirrors is weak. We can linearize
the fluctuations around the steady state using Eqs.(\ref{langevin}) and
obtain a set of linear quantum Langevin equations for the fluctuation
operators. We define $f=(\delta Q_{1},\delta P_{1,}\delta Q_{2},\delta
P_{2},\delta X_{1},\delta Y_{1},\delta X_{2},\delta Y_{2},\delta
U_{1},\delta U_{2},\delta V_{1},\delta V_{2})^{T}$. A set of linear quantum
Langevin can be written as
\begin{equation}
{\dot{f}}=Af+B  \label{linear}
\end{equation}%
where
\begin{widetext}
\begin{equation*}
A=\left[
\begin{array}{cccccccccccc}
0 & \omega _{m_{1}} & 0 & 0 & 0 & 0 & 0 & 0 & 0 & 0 & 0 & 0 \\
-\omega _{m_{1}} & -\gamma _{m_{1}} & 0 & 0 & -\chi _{1}X_{1}^{s} & -\chi
_{1}Y_{1}^{s} & 0 & 0 & 0 & 0 & 0 & 0 \\
0 & 0 & 0 & \omega _{m_{2}} & 0 & 0 & 0 & 0 & 0 & 0 & 0 & 0 \\
0 & 0 & -\omega _{m_{2}} & -\gamma _{m2} & 0 & 0 & -\chi _{2}X_{2}^{s} &
-\chi _{2}Y_{2}^{s} & 0 & 0 & 0 & 0 \\
\chi _{1}Y_{1}^{s} & 0 & 0 & 0 & -\kappa _{1} & \widetilde{\Delta _{1}} & 0
& 0 & 0 & g_{1} & 0 & 0 \\
-\chi _{1}X_{1}^{s} & 0 & 0 & 0 & -\widetilde{\Delta _{1}} & -\kappa _{1} & 0
& 0 & -g_{1} & 0 & 0 & 0 \\
0 & 0 & \chi _{2}Y_{2}^{s} & 0 & 0 & 0 & -\kappa _{2} & \widetilde{\Delta
_{2}} & 0 & 0 & 0 & g_{2} \\
0 & 0 & -\chi _{2}X_{2}^{s} & 0 & 0 & 0 & -\widetilde{\Delta _{2}} & -\kappa
_{2} & 0 & 0 & -g_{2} & 0 \\
0 & 0 & 0 & 0 & 0 & -g_{1}r_{a}\rho _{aa}^{0} & 0 & g_{2}r_{a}\rho _{ac}^{0}
& -\gamma & \delta _{1} & 0 & 0 \\
0 & 0 & 0 & 0 & g_{1}r_{a}\rho _{aa}^{0} & 0 & g_{2}r_{a}\rho _{ac}^{0} & 0
& -\delta _{1} & -\gamma & 0 & 0 \\
0 & 0 & 0 & 0 & 0 & -g_{1}r_{a}\rho _{ca}^{0} & 0 & g_{2}r_{a}\rho _{cc}^{0}
& 0 & 0 & -\gamma & \delta _{2} \\
0 & 0 & 0 & 0 & -g_{1}r_{a}\rho _{ca}^{0} & 0 & -g_{2}r_{a}\rho _{cc}^{0} & 0
& 0 & 0 & -\delta _{2} & -\gamma%
\end{array}%
\right]
\end{equation*}%
\end{widetext}, $B=(0,\xi _{1},0,\xi _{2},\sqrt{2\kappa _{1}}\delta X_{1in},%
\sqrt{2\kappa _{1}}\delta Y_{1in},\sqrt{2\kappa _{2}}\delta X_{2in}$, $\sqrt{%
2\kappa _{2}}\delta Y_{2in},0,0,0,0)^{T}$. We have defined
\begin{eqnarray}
X_{j} &=&\frac{1}{\sqrt{2}}(a_{j}+a_{j}^{\dagger }),Y_{j}=\frac{1}{\sqrt{2}i}%
(a_{j}-a_{j}^{\dagger }),j=1,2, \\
U_{1} &=&\frac{1}{\sqrt{2}}(\sigma _{ba}+\sigma _{ab}),U_{2}=\frac{1}{\sqrt{2%
}i}(\sigma _{ba}-\sigma _{ab}),  \notag \\
V_{1} &=&\frac{1}{\sqrt{2}}(\sigma _{ba}+\sigma _{ab}),V_{2}=\frac{1}{\sqrt{2%
}i}(\sigma _{cb}-\sigma _{bc}).  \notag
\end{eqnarray}%
A is the drift matrix. The system is stable only if the real part of all the
eigenvalues of the matrix A are negative, which is also the requirement of
the validity of the linearization method. Because of the atomic cascade
form, the system can be a amplifying system, that is to say, for some
parameters, the eigenvalues of drift matrix A can have a positive real part.
So, we must carefully eliminate the parameters region in order to retain it
within the stability conditions. Due to the twelve dimensions of drift
matrix A, it is not easy to obtain the analytic expression of the
requirement as \cite{Genes} eigenvalues. We will guarantee the requirement
via numerical method. All the parameters chosen in this paper have been
verified to satisfy the stability conditions.

\subsection{Entanglement of the nanomechanical oscillators and the two-mode
fields within the cavity}

We investigate the nature of linear quantum correlations among fields and
mirrors by considering the steady state of the correlation matrix of quantum
fluctuations in this multipartite system. The quantum noises $\xi $ and $%
a_{i}^{in}$ are zero-mean quantum Gaussian noises and the dynamics has been
linearized, as a consequence, the steady state of the system is a zero-mean
multi-partite Gaussian state. We defined $V_{ij}(\infty )=\frac{1}{2}%
[\langle f_{i}(\infty )f_{j}(\infty )+f_{j}(\infty )f_{i}(\infty )\rangle ]$%
, the element of covariance matrix. One can obtain various correlation
information from it, and we will show the entanglement of the two mirrors as
well as the two-mode fields after obtaining the covariance matrix.

We now recall entanglement criteria of continuous variable proposed by Duan %
\cite{duan} and Simon \cite{simon2}. According to \cite{duan}, a state is
entangled if the summation of the quantum fluctuations in the two EPR-like
operators $X$ and $Y$ satisfy the following inequality
\begin{equation}
(\Delta X)^{2}+(\Delta Y)^{2}<2.
\end{equation}%
For the mechanical oscillator we define $X_{m}=Q_{1}+Q_{2},Y_{m}=P_{1}-P_{2}$
while for the two mode fields $X_{f}=X_{1}-X_{2},Y_{f}=Y_{1}+Y_{2}$. The
criterion in Eq.(11) can be directly detected in experiment via homodyne
measurements \cite{villar}. Simon's criterion has more direct relation with
covariance matrix proposed in \cite{simon2} and developed in \cite{gaussian}%
. For a physical state, the covariance matrix must obey the
Robertson-Schrodinger uncertainty principle,%
\begin{equation}
V+\frac{i}{2}\beta \geq 0,  \label{uncertainty}
\end{equation}%
where $\beta =\left[
\begin{array}{cc}
J & 0\\
0 & J%
\end{array}
\right] $ with $J=\left[
\begin{array}{cc}
0& 1\\
-1 & 0%
\end{array}
\right]$ when we define vector $f=(\delta q_1,\delta p_1,\delta q_2,\delta p_2)^T$ for two-mode fields \cite{simon2}. If a state is separable, partial
transpose matrix $\widetilde{V}$ (obtained from $V$ just by taking $p_{j}$
in $-p_{j}$ ) still obey the inequality in (\ref{uncertainty}). The
inequality requires that all the symplectic eigenvalues of the transposed
matrix are larger than 1/2 in terms of the definition of Eq.(10). The
symplectic eigenvalues can be calculated from the square roots of the
ordinary eigenvalues of $-\left( \beta \widetilde{V}\right) ^{2}$\cite%
{CASSE,zhao}. So, if the smallest eigenvalue is smaller than 1/2, the
transposed mode is then inseparable. For the two-mode Gaussian states, the violation of the inequality is a
sufficient and necessary condition for the existence of entanglement between the
transposed mode and the remaining modes.

Neglecting the frequency dependence as it was pointed in \cite{Genes}, under
the condition of Markovian approximation, the frequency domain treatment is
equivalent to the time domain derivation considered in \cite{genes2}. When
the stability conditions (all real parts of eigenvalues of matrix A are
negative) are satisfied, the steady state correlation matrix of the quantum
fluctuation meet the following Lyapunov equation \cite{Genes}%
\begin{equation}
AV+VA^{T}=-D,
\end{equation}%
where $D=Diag[0,\gamma _{m_{1}}(2N+1),0,\gamma _{m_{1}}(2N+1),\kappa
_{1}(2N+1),\kappa _{1}(2N+1),\kappa _{2}(2N+1)$, $\kappa _{2}(2N+1),0,0,0,0]$%
.

\begin{figure}[b]
\includegraphics[width=\columnwidth]{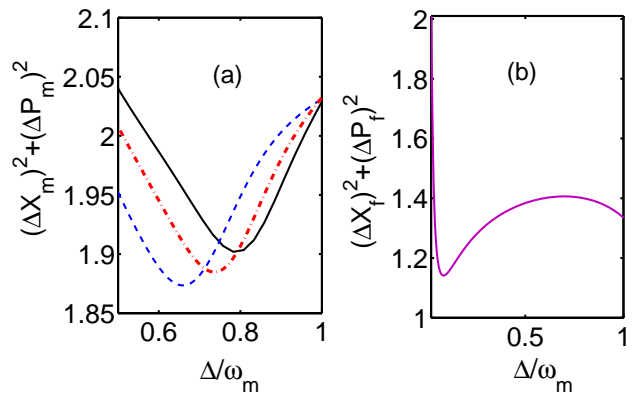}
\caption{(Color online) (a) shows the entanglement of the mechanical mirror
where $g_{1}=2\protect\pi \times 1.5\times 10^{5}Hz$ (solid), $2\protect\pi %
\times 1.7\times 10^{5}Hz$ (dash-dot), $2\protect\pi \times 2.0\times
10^{5}Hz$ (dashed). (b) plot the entanglement of the two mode fields where $%
g1=2\protect\pi \times 2.2\times 10^{5}Hz$ . For all the plots, the other
parameters are $L=5mm$, $m=20ng$, $\protect\kappa _{1}=\protect\kappa _{2}=2%
\protect\pi \times 215KHz$, $\protect\omega _{m_{1}}=\protect\omega %
_{m_{2}}=2\protect\pi \times 10MHz$. The wavelength of the laser $\protect%
\lambda =810nm$ with power $10mW$, and the mechanical quality factor $%
Q\prime =\frac{\protect\omega _{m}}{\protect\gamma _{m}}=6700$, $ra=2000$, $%
r=1.3MHz$, $\protect\delta _{1}=\protect\delta _{2}=4MHz$, $T=42\protect\mu K
$ . }
\end{figure}

\begin{figure}[b]
\includegraphics[width=\columnwidth]{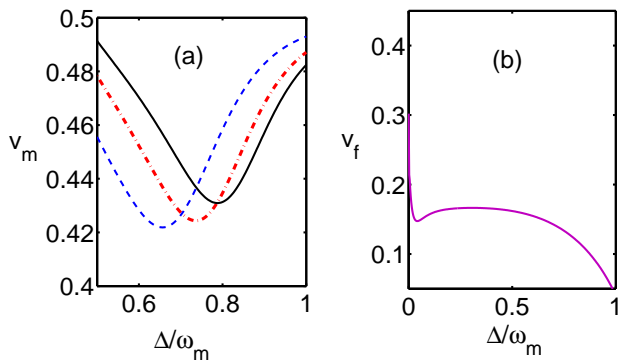}
\caption{(Color online) The entanglement of the mechanical mirrors and
two-mode fields via the smallest eigenvalues of partial transposed matrix.
All the lines and parameters are correspond to Fig.2.}
\end{figure}

For simplicity, we choose all parameters of the two mirrors and the two mode
cavities are the same, for example $g_{1}=g_{2}$, $\omega _{m_{1}}=\omega
_{m_{2}}$, $\omega _{L_{1}}=\omega _{L_{2}}$, etc.. Then the effective
detuning $\widetilde{\Delta }_{1}=\widetilde{\Delta }_{2}=\Delta $. In
Fig.2, we choose the atoms injected into the cavity with initial state $%
|\Psi _{a}(0)\rangle =\frac{1}{\sqrt{2}}(|a\rangle +|c\rangle )$ which is
the best coherence of the two levels. We numerically show the entanglement
of the movable mirrors as well as the two mode fields induced by atomic
coherence. Fig. 2a show that with the increasing of the coupling between
cavity and the atoms, the deep squeezing in fluctuation move towards the
side of small value of $\Delta $. As already shown in \cite{Genes,marq},
many features of the radiation-pressure interaction in the cavity can be
understood by considering that the driving laser light is scattered by the
vibrating cavity boundary mostly at the first Stokes $\omega _{0}-\omega
_{m} $ and anti-Stokes $\omega _{0}+\omega _{m}$. Therefore, the
optomechanical interaction will be enhanced when the cavity is resonant with
one of the two sidebands, i.e., when $\Delta =\omega _{m}$, where
entanglement between cavity field and mirror is enhanced. Here, due to the
atomic medium existence, the two sideband move towards its middle. In
addition, within the region of parameter in Fig. 2a we can observe that the larger value of the coupling,
the deeper squeezing in\ the fluctuation. Undoubtedly, the entanglement is
resulted from the interaction between atoms and the fields as well as the
atomic level coherence, therefore, the larger coupling $g_{i}$ the deeper
squeezing within our parameters. The enhancement of $g_{i}$ can be substituted by increasing the
injection rate of the atoms. Fig. 2b show that the two-mode fields are
entangled for nearly the whole region of detuning $\Delta $ (from 0 to 1).
Using the same group of parameters, from the solution of Eq.(13) we reconstruct two-partite
covariance matrixes for mirrors and two-mode fields and show their smallest
eigenvalues under partial transpose, shown in Fig.3. One can see that Fig.3
are almost the same in shape with Fig.2, that is to say, the entanglement
judged with two criteria are equal. By showing the entanglement with two
criteria, we prove our correction of calculation and convince the
entanglement existence.

For cascade atomic configuration system, the process of radiation two-mode
fields are correspondence to parametric-down-conversion process. So, one can
easy understand the two-mode entanglement. For fixed mirror, it has been
known that two cavity modes are entangled when they interact with cascade
three-level atoms \cite{han}. Here, we show further that when one mirror is
movable, the two-mode entangled continuous-variable state still hold. In our
numerical simulation, we find that there is a threshold coupling between
atoms and fields (under these group of parameters), below that we have no
two entangled mirrors, which means that the appropriate coupling between
atoms and cavity is favorable if we would have entangled mirrors. Thus, the
radiation fields emitted from cascade atoms can be entangled, and the
entanglement of the two-mode field can be transferred into the mechanical
mirror because of the radiation-pressure interaction. As a consequence, we
have both the entanglement of two-mode fields and two mirrors.

\subsection{The output two-mode field entanglement with optomechanical
oscillator}

Because one can not direct access to the intracavity fields, if one would
like to use the entanglement, he need to export the entangled fiedls first; thus
only the output entanglement of the optical cavity has practical meaning .
In addition, by means of spectral filters, one can always select many
different traveling output modes originating from a single intracavity mode,
which offer us the opportunity to easily produce and manipulate a
multipartite system, eventually possessing multipartite entanglement \cite%
{Genes}. So, in this section we study the entanglement of the output field.
The input-output relation for the two mode fields are
\begin{equation}
\delta a_{jout}=\sqrt{2\kappa _{j}}\delta a_{j}-\delta a_{jin},j=1,2.
\label{outa}
\end{equation}%
The linear equation of (\ref{linear}) can be solved in the frequency domain
by Fourier transformation with the solution $f(\omega )=(-i\omega -A)^{-1}B$%
. In the interaction picture $\omega $ represents the detuning from the
cavity frequency \cite{barberis}. Considering the relation of (\ref{outa}),
we have
\begin{equation}
f^{out}(\omega )=C(-i\omega -A)^{-1}B-E,
\end{equation}%
where $C=diag(0,0,0,0,\sqrt{2\kappa _{1}},\sqrt{2\kappa _{1}},\sqrt{2\kappa
_{2}},\sqrt{2\kappa _{2}},0,0,0,0)$, $E=[0,0,0,0,\delta X_{1in}(\omega
),\delta Y_{1in}(\omega ),\delta X_{2in}(\omega ),\delta Y_{2in}(\omega ),0,0
$, $0,0]^{T}$. We can finally obtain the output correlation matrix $%
V_{ij}^{out}(\omega )=\frac{1}{2}[\langle f_{i}^{out}(\omega
)f_{j}^{out}(\omega ^{\prime })+f_{j}^{out}(\omega ^{\prime
})f_{i}^{out}(\omega )\rangle ]$, and the squeezing spectrum
\begin{eqnarray}
S_{OUT}(\omega ) &=&\frac{1}{2}[\delta X_{f}(\omega )\delta X_{f}(\omega
^{\prime })+\delta X_{f}(\omega ^{\prime })\delta X_{f}(\omega )  \notag \\
&&+\delta Y_{f}(\omega )\delta Y_{f}(\omega ^{\prime })+\delta Y_{f}(\omega
^{\prime })\delta Y_{f}(\omega ^{\prime })]
\end{eqnarray}%
can be calculated from the correlation matrix. The spectrum is defined in a
frame of \ \
\begin{figure}[b]
\includegraphics[width=\columnwidth]{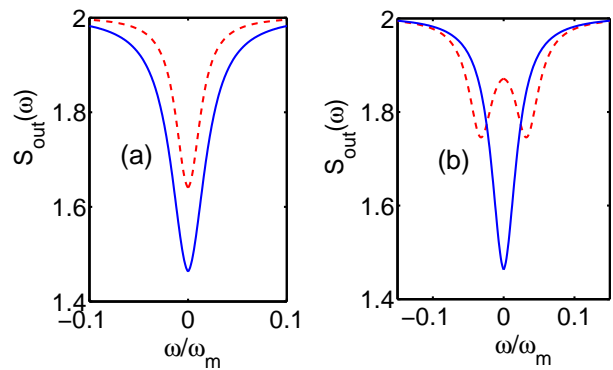}
\caption{(Color online) (a) polts the output squeezing spectrum for several
values of frequency of the mirrors where $\protect\omega _{m_{1}}=\protect%
\omega _{m_{2}}=2\protect\pi \times 10MHZ$ (solid) $2\protect\pi \times 15MHZ
$ (dashed). (b) shows the squeezing spectrum under different input-atomic
state . For solid line $\protect\rho _{aa}=\protect\rho _{cc}=\protect\rho %
_{ac}=0.5$, for dashed line $\protect\rho _{aa}=\frac{1}{5}$, $\protect\rho %
_{cc}=\frac{4}{5}$, $\protect\rho _{ac}=\frac{2}{5}.$ In Fig.4b, $\protect%
\omega _{m_{1}}=\protect\omega _{m_{2}}=2\protect\pi \times 10MHZ$. For all
the line in Fig.4, $g1=g_{2}=2\protect\pi \times 2.0\times 10^{5}Hz$ , $%
\Delta =0.8\protect\omega _{m_{1}}$, and the other parameters are the same
with Fig.2.}
\end{figure}

Fig.4 shows the two-mode output squeezing spectrum, which can also be
considered as a measure of entanglement in frequency domain\cite{zhoupla}.
In Fig.4a, we notice that deep squeezing (usually means that the maximum
entanglement) happen when $\omega =0$, that is to say, when the output
frequency is equal to that of cavity frequency, we have maximum
entanglement. It is very interesting to observe that the deep squeezing are
affected by the frequency of optomechanical oscillator. The larger frequency
of the mirror, the smaller entanglement. We can deduce that if the mirrors
is fixed, we should have the deepest squeezing in fluctuation of the output
fields. The more higher frequency of the mirrors, the more entanglement can
be transferred from the fields to the mirrors so that the smaller
entanglement is leaved in the output fields.

Fig.4b shows the squeezing spectrum affected by the initial atomic state. As
we have discuss above, the entanglement of two-mode fields as well as the
two mirrors originate from the atomic coherence. Undoubtedly, the initial
atomic coherence affects both the intracavity entanglement and the output
entanglement. For the best coherence of atomic initial state $\rho _{a}^{0}=%
\frac{1}{2}(|a\rangle +|c\rangle )(\langle a|+\langle c|)$, we have one
valley of deep squeezing when the output frequency is the same as to that of
cavity ($\omega =0$ in Fig. 4a). But for the input atomic state $\rho
_{a}^{0}=\frac{1}{5}|a\rangle \langle a|+\frac{2}{5}(|a\rangle \langle
c|+|c\rangle \langle a|)+\frac{4}{5}|c\rangle \langle c|$, the maximum
squeezing have two valley. As to the split of the squeezing spectrum, we can
understand it from initial atomic state. The initial population in $%
|a\rangle $ and $|c\rangle $ are no longer equal so that the average of
steady state $|X_{i}^{s}|$ ($|Y_{i}^{s}|$) for the two modes are not yet
equal. Then, the fluctuation the $\delta X_{1}$ ($\delta Y_{1}$) and $\delta
X_{2}$ ($\delta Y_{2}$) can be different so that the squeezing spectrum of
the product of the fluctuation change from one degeneracy valley into two
maximum squeezing. The split of the squeezing spectrum from one valley into
two minima also have been shown in the case of asymmetric loss for each
mode, i.e., $\kappa _{1}\neq \kappa _{2}$ \cite{walls}. Our group had
revealed the similar split resulting from asymmetric detuning and asymmetric
atomic initial state \cite{zhoupla}.

During above numerical simulation, we choose parameters of the system based
on recent experiment for optomechanical system \cite{simon}, also
considering the parameters used in theoretical papers \cite%
{Genes,huangshumei}. As to the coupling between the atoms and the cavity, we
use $g=g1=g_{2}=2\pi \times 2.8\times 10^{5}Hz$ ( in Fig.4). Comparing with
recent experiment $g/\pi =50MHz$ \cite{coupling}, the coupling is much more
weak. Moreover, the ratio $g/\kappa $ is only $1.3$, see Fig.4, which can be
realized in recent experiment technique \cite{fabry}.

\section{Conclusion}

We proposed a scheme via three-level cascade atom to entangle two-mode
fields as well as optomechanical oscillator. Our study show that intracavity
fields and two movable mirrors are entangled respectively for realizable
coupling between the cavity fields and the atoms. The output two-mode fields
entanglement is affected by the frequency of the mirror motion, the larger
frequency of the mirror, the larger entanglement of the output fields.
Because the entanglement is resulted from the atomic coherence, the initial
atomic state play important role on the output entanglement of the two-mode
fields. For the best coherence of the initial atomic state, the output
spectrum appears one deep squeezing while for other initial atomic state,
the one deep squeezing split into two deep squeezing.

\begin{acknowledgments}
Acknowledgments: L.Z.and Y.H. are supported by NSFC (Grant
Nos.10774020,11074028). Jietai Jing is supported by NSFC ( Grant No
10974057),Shanghai Pujiang Program (Grant No.09PJ1404400), and the program
for Professor of Special Appointment (Eastern Scholar) at Shanghai
Institutions of higher Learning. Weiping Zhang acknowledges the support of
NSFC (Grant Nos. 10828408, 10588402), the National Basic Research Program of
China (973 Program) under Grant No. 2011CB921604. All authors thank Open
Fund of the State Key Laboratory of Precision Spectroscopy, ECNU.
\end{acknowledgments}

\end{document}